\documentclass[letterpaper,11pt,reqno,tbtags]{amsart} 
\usepackage[margin=3cm]{geometry} 
\usepackage{mathrsfs}
\usepackage{amsmath}
\usepackage[
colorlinks=true,%
breaklinks,
hyperindex,%
plainpages=false,%
bookmarksopen=false,%
linkcolor=blue,%
anchorcolor=blue,%
citecolor=blue,%
filecolor=black,%
menucolor=black,%
pagecolor=black,%
urlcolor=blue,%
pdfview=FitH,
pdfstartview=FitH,
hyperfigures,
bookmarksnumbered%
]{hyperref} 
\usepackage{amsmath,amssymb,amsthm,amsfonts,amsbsy,latexsym,dsfont,tikz,multirow}
\usepackage[skip=5pt]{caption}

\usepackage{lipsum,afterpage,refcount}
\newcommand{\setfootnotemark}{%
  \refstepcounter{footnote}%
  \footnotemark[\value{footnote}]}

\usepackage{algorithm2e}
\usepackage{enumerate}

\newcommand{\matnorm}[1]{{\left\vert\kern-0.25ex\left\vert\kern-0.25ex\left\vert #1 
\right\vert\kern-0.25ex\right\vert\kern-0.25ex\right\vert}}

\newcommand{\mbf}[1]{\mbox{\boldmath$#1$}}

\newcommand{\NPI}{\mathsf{NPI}}

\theoremstyle{remark}

\renewcommand{\leq}{\leqslant}

\newcommand{\rd}{\mathrm{d}}

\newcommand{\cR}{\mathcal{R}}

\newcommand{\vone}{\mathbf{1}}

\newcommand{\bR}{\mathbb{R}}

\DeclareMathOperator{\argmin}{argmin}

\begin{document}

\title[Impact of interventions on COVID-19]{Scenario analysis of non-pharmaceutical interventions on global COVID-19 transmissions}

\author[Xiaohui Chen]{Xiaohui Chen}

\author[Ziyi Qiu]{Ziyi Qiu}

\date{First arXiv version: April 7, 2020. This version: \today}
\thanks{Xiaohui Chen was with Department of Statistics at University of Illinois at Urbana-Champaign and Institute for Data, System, and Society (IDSS) at Massachusetts Institute of Technology. {\it Email:} xiaohui@mit.edu. Ziyi Qiu was with Department of Economics at University of Illinois at Urbana-Champaign and Department of Economics at University of Chicago. {\it Email:} ziyiqiu@illinois.edu. Research was supported in part by NSF CAREER Award DMS-1752614 and a Simons Fellowship.}

\begin{abstract}
This paper introduces a dynamic panel SIR (DP-SIR) model to investigate the impact of non-pharmaceutical interventions (NPIs) on the COVID-19 transmission dynamics with panel data from 9 countries across the globe. By constructing scenarios with different combinations of NPIs, our empirical findings suggest that countries may avoid the lockdown policy with imposing school closure, mask wearing and centralized quarantine to reach similar outcomes on controlling the COVID-19 infection. Our results also suggest that, as of April 4th, 2020, certain countries such as the U.S. and Singapore may require additional measures of NPIs in order to control disease transmissions more effectively, while other countries may cautiously consider to gradually lift some NPIs to mitigate the costs to the overall economy.
\end{abstract}

\maketitle

\section{Introduction} 
Since December 2019, a coronavirus disease (COVID-19) has been spreading in China and now emerging as a global pandemic. In response to the COVID-19 crisis, many countries have ordered unprecedented non-pharmaceutical interventions (NPIs), including travel restriction, mask wearing, lockdown, social distancing, school closure, and centralized quarantine (such as cabin hospitals), all aiming to reduce the population contact rates and thus mitigate the virus transmission. On the other hand, prolonged NPIs have a large downside impact on the overall economic and social well-being, which may cause major concerns including increased unemployment rates and bankruptcy of firms. Therefore, it is of the utmost importance to investigate which interventions and to what extend have substantial impact on controlling the epidemiological dynamics, and how to choose the appropriate measures to fit country-specific socioeconomic circumstance. Understanding the impact of various NPIs on a global scale can provide insights for each country to choose the most cost-effective NPIs in a timely manner to contain and mitigate the virus spread.

A number of recent works studied the government intervention effects. \cite{ChuNgLin2020_Lancet} applied the Bass Susceptible-Infected-Recovered (Bass-SIR) model to study the lockdown and social distancing effects for province-specific epidemiological parameters in China. \cite{Flaxman2020_NPIs} used only the observed death data and proposed a (non-SIR based) Bayesian model to study several intervention effects on 11 European countries. \cite{Ferguson2020_NPI} modified an individual-based simulation model to study the consequence of NPIs to reduce the COVID-19 mortality and healthcare demand in the UK and the U.S. \cite{Wang2020} developed a Susceptible-Exposed-Infectious-Removed (SEIR) model to evaluated the impact of NPIs on the epidemic in Wuhan, China. \cite{Laietal2020} built a travel network-based SEIR model to study the impact of different NPIs in China. \cite{Chenetal2020} proposed a time-dependent SIR model to account the impact of the Wuhan city lockdown and predicted the future trend of the COVID-19 transmission. 

During the COVID-19 transmission process across the globe, several countries experienced an epidemic outbreak in 2020 at different timelines. The first COVID-19 epidemic outbreak started in Wuhan, China in late January. In early to mid-February, Singapore was near the top list of total confirmed cases outside China, and South Korea began to see rising numbers of total cases. A few days after the outbreak in South Korea in mid-February, outbreaks in Iran and Italy began in late February. In early March, the number of cases in many European countries sharply increased, shifting the epicenter from Asia to Europe. As of April 4th, European countries with the most confirmed cases include Italy, Spain, France, Germany, and the UK. While China and South Korea managed to control the virus spread and European countries are experiencing the outbreak, the U.S. quickly emerged as the biggest epicenter in the world, with the total cases surpassed China and Italy in late March, becoming the country with the most confirmed cases currently.

While countries experienced different timelines of the COVID-19 epidemic outbreaks, the non-pharmaceutical interventions each government imposed also varied accordingly. China implemented very strict NPIs right after the outbreak occurred in Wuhan, including travel restriction~\cite{china_TR}, lockdown~\cite{china_LD}, school closure~\cite{china_SC}, social distancing~\cite{china_SD}, wearing masks~\cite{china_MW}, and later on centralized quarantine~\cite{china_CQ}. South Korea imposed travel restriction~\cite{sk_TR}, school closure~\cite{sk_SC}, social distancing~\cite{sk_SD}, wearing masks~\cite{sk_MW}, and centralized quarantine~\cite{sk_CQ}, however it did not call a national lockdown. Singapore implemented travel restriction~\cite{sg_TR}, social distancing~\cite{sg_SD}, and centralized quarantine~\cite{sg_CQ}, without emphasis on mask wearing, school closure and a national lockdown. European countries have very similar government policies (at different time points) including travel restrictions, social distancing, school closure, and lockdown~\cite{it_TR,it_LD,it_SD,it_SC,esp_TR,esp_LD,esp_SD,esp_SC,fr_TR,fr_LD,fr_SD,fr_SC,ge_TR,ge_LD,ge_SD,ge_SC,uk_TR,uk_LD,uk_SD,uk_SC} but not imposing policies on wearing masks and centralized quarantine. The U.S. has closer policies relative to Europe~\cite{us_TR,us_SD,us_SC}, except that it did not call a national lockdown yet (as of April 4th, 2020).

Due to the time-varying and heterogeneous nature of the outbreaks and the associated different NPIs across countries in the COVID-19 pandemic, it is important to borrow information from the panel data collected worldwide to help understand the impacts of different NPIs and to improve the prediction of the epidemiological developments for countries at later stages. To the best of our knowledge, there is no up-to-date analysis based on integrated infection, recovery and death data from different countries with significant variations in their NPIs and timelines on the global scale. In this paper, we perform a scenario analysis of NPIs on the COVID-19 transmission through a dynamic SIR model tailored to a panel study with data from 9 countries in three continents, which were or are currently the epicenters: Italy, Spain, Germany, France, the UK, Singapore, South Korea, China, and the United States. 


\section{Methodology}
\label{sec:model}

\subsection{SIR model with time-varying parameters}
\label{subsec:sir}

The SIR model is a fundamental compartmental model in epidemiology~\cite{Hethcote2000_SIAM}. The dynamic hypothesis behind the SIR model is the Kermack-McKendrick theory that predicts the number of cases of an infectious disease as it is transmitted through a population over time. In this paper, we consider a time-varying SIR model described by the following system of (non-linear) ordinary differential equations (ODEs):
\begin{eqnarray*}
{\rd S \over \rd t} &=& - {\beta(t) I(t) \over N} S(t) , \\
{\rd I \over \rd t} &=& {\beta(t) S(t) \over N} I(t) - \gamma(t) I(t), \\
{\rd R \over \rd t} &=& \gamma(t) I(t),
\end{eqnarray*}
where $\beta(t) > 0$ is the disease transmission rate at time $t$ and $\gamma(t) > 0$ is the recovery rate at time $t$. Here $S(t), I(t), R(t)$ denote the subpopulation sizes of susceptible, infected, and recovered at time $t$, respectively. In the absence of vaccine, we assume that the whole population size $N = S(t) + I(t) + R(t)$ is constant over time. Similar time-varying SEIR model (with an extra exposure state) was considered in~\cite{LekoneFinkenstadt2006_Biometrics}. The time-varying {\it reproduction number} is defined as
\begin{equation}
\label{eqn:reproduction_number}
\cR_{t} = {\beta(t) \over \gamma(t)}.
\end{equation}
In particular for $t = 0$, $\cR_{0}$ is the {\it basic reproduction number} and $\cR_{\mbox{eff}} = \cR_{0} {S(0) \over N}$ is the {\it effective reproduction number}. If $\cR_{\mbox{eff}} > 1$, then $I(t)$ will first increase to its maximum and then decrease to zero, which is often referred as an epidemic outbreak. If $\cR_{\mbox{eff}} < 1$, then $I(t)$ is monotonically decreasing and there would be no epidemic outbreak.

\subsection{Dynamic panel SIR model}
\label{subsec:dp-sir}

Suppose we have collected a panel dataset (i.e., longitudinal data) from $p$ countries of population size $N_{1},\dots,N_{p}$, and the data points are observed on evenly spaced time intervals such as daily data. Let $S_{j}(t), I_{j}(t), R_{j}(t)$ be the subpopulation sizes of susceptible, infected, recovered in country $j$ at time $t$, respectively. Given initial data $\{ S_{j}(0), I_{j}(0), R_{j}(0) \}_{j=1}^{p}$, the Euler method with the unit time interval gives the discretized version of the following {\it dynamic panel SIR} (DP-SIR) model determined by the system of equations: for $j = 1, \dots,p$,
\begin{eqnarray*}
I_{j}(t+1) - I_{j}(t) &=& {\beta_{j}(t) S_{j}(t) \over N_{j}} I_{j}(t) - \gamma_{j}(t) I_{j}(t), \\
R_{j}(t+1) - R_{j}(t) &=& \gamma_{j}(t) I_{j}(t), \\
N_{j} &=& S_{j}(t) + I_{j}(t) + R_{j}(t),
\end{eqnarray*}
where $\beta_{j}(t) > 0$ and $\gamma_{j}(t) > 0$ are the rates of disease transmission and recovery in country $j$ at time $t$, respectively. Denote $\Delta I_{j}(t) = I_{j}(t+1)-I_{j}(t)$ and $\Delta R_{j}(t) = R_{j}(t+1)-R_{j}(t)$. Then we may rewrite the discrete DP-SIR model as
\begin{eqnarray*}
\Delta I_{j}(t) + \Delta R_{j}(t) &=& {\beta_{j}(t) S_{j}(t) \over N_{j}} I_{j}(t), \\
\Delta R_{j}(t) &=& \gamma_{j}(t) I_{j}(t),
\end{eqnarray*}
subject to the constraint $N_{j} = S_{j}(t) + I_{j}(t) + R_{j}(t)$. For simplicity, we treat $\gamma_{j}(t), j=1,\dots,p$ as constant functions, i.e., $\gamma_{j} = \gamma_{j}(t)$. To model the intervention effect on $\beta(t)$, our basic intuition is that each intervention has the same effect on the disease transmission rate (thus the reproduction numbers since $\gamma_{j}(\cdot)$ is constant) across countries and over time. Specifically, to borrow information cross countries, the coefficients $\beta_{j}(t)$ have a panel data structure:
\begin{equation}
\label{eqn:panel_structure}
\beta_{j}(t) = \exp \Big( \alpha_{j} + \sum_{k=1}^{K} \beta_{k} \; \NPI_{tjk} \Big),
\end{equation}
where $\NPI_{tjk}$ is the $k$-th NPI in country $j$ at time $t$, $\alpha_{j}$ is the country-level fixed (non-random) effect, and $\beta_{k}$ is the $k$-th NPI effect which does not depend on $j$ and $t$.

\subsection{Construction of the NPIs}
\label{subsec:modeling_NPIs}

We now construct the NPIs from their intervention times. Let $t \in [0, T]$ be the time span for observing data from the DP-SIR model and $t^*_{jk}$ be the intervention time by the $k$-th NPI in country $j$. Then $\NPI_{tjk}$ is modeled as following:
\begin{equation}
\label{eqn:model_NPIs}
\NPI_{tjk} = \left\{
\begin{array}{cc}
1 & \mbox{if } 0 \leq t < t^*_{jk} \\
\exp \Big(-{t-t^*_{jk} \over \tau} \Big) & \mbox{if } t^*_{jk} \leq t \leq T \\
\end{array}
\right. ,
\end{equation}
where $\tau > 0$ is a user-specified scale parameter controlling the time-lag effect of interventions. Specifically, for smaller $\tau$, $\NPI_{tjk}$ is closer to the indicator function $\vone(t \leq t^*_{jk})$; for larger $\tau$, $\NPI_{tjk}$ decays to zero more slowly, which reflects the time-lag to see the intervention effect. Thus $\NPI_{tjk}$ in~\eqref{eqn:model_NPIs} is a smooth approximation of interventions with incorporated time-lag effect.

\subsection{Estimation}
\label{subsec:estimation}

Suppose the data $I_{j}(t)$ and $R_{j}(t)$ are observed at time points $t_{i}, i =1,\dots,T_{j}$ with equal spacing. With the panel structure~\eqref{eqn:panel_structure} and the NPI parametrization~\eqref{eqn:model_NPIs}, our goal is to estimate the parameter of interest $\mbf\theta := (\alpha_{1},\dots,\alpha_{p}, \beta_{1},\dots,\beta_{K}, \gamma_{1}, \dots, \gamma_{p}) \in \bR^{K+2p}$ from the observed data. For shorthand notation, we write $I_{ij} = I_{j}(t_{i}), \Delta I_{ij} = I_{j}(t_{i}+1) - I_{j}(t_{i})$ for $i = 1,\dots,T_{j}-1$, and similarly for $R_{ij}, S_{ij}$ and $\Delta R_{ij}, \Delta S_{ij}$. Denote $\NPI_{ijk} = \NPI_{t_{i}jk}$.

We use the ordinary least-squares (OLS) to estimate $\mbf\theta$. The squared loss function of the DP-SIR model is given by
\begin{equation*}
\ell(\mbf\theta) = \sum_{j=1}^{p} \sum_{i=1}^{T_{j}-1} \Big[ \log \Big( \Delta I_{ij} + \Delta R_{ij} \Big) - \log \Big( {S_{ij} I_{ij} \over N_{j}} \Big) -\Big( \sum_{k=1}^{K} \beta_{k} \; \NPI_{ijk} + \alpha_{j} \Big) \Big]^{2} + \sum_{j=1}^{p} \sum_{i=1}^{T_{j}-1} \Big( \Delta R_{ij} - \gamma_{j} I_{ij} \Big)^{2}.
\end{equation*}
Then the OLS estimate for $\mbf\theta$ is given by
\begin{equation*}
\hat{\mbf\theta} = \argmin_{\mbf\theta \in \bR^{K+2p}} \ell(\mbf\theta).
\end{equation*}
Based on the OLS estimate, we can also test for which intervention is statistically significant.

\section{Scenario analysis on COVID-19 panel data}
\label{sec:covid-19_data}

\subsection{Data description}
\label{subsec:data_description}

\begin{figure}[h]
\centering
\caption{Observed data for the number of active cases from 9 countries between 1/22/20 and 4/3/20. There are $T = 73$ data points in each plot.}
\label{fig:observed_data_COVID-19}
\includegraphics[scale=0.7]{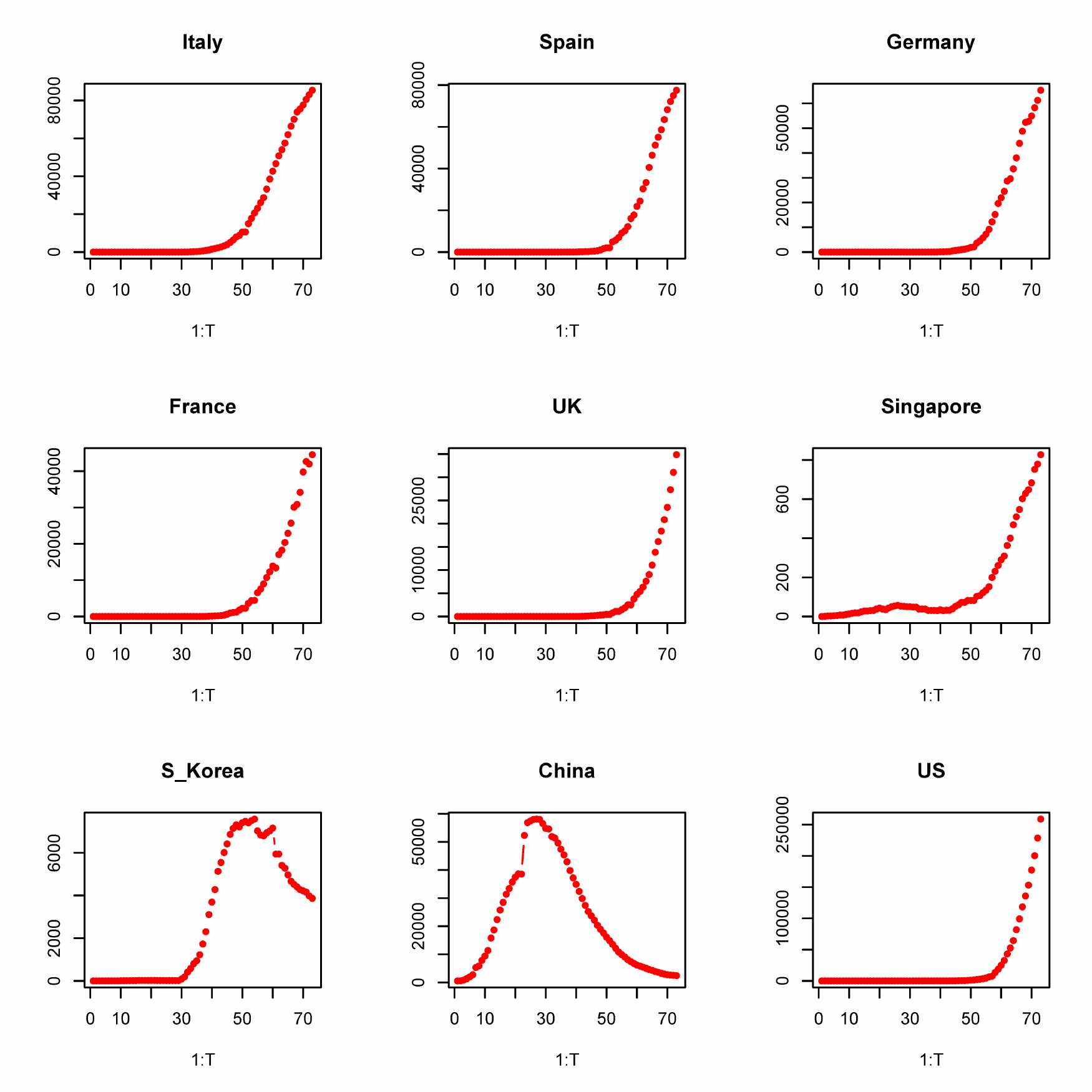}
\end{figure}

We collected COVID-19 data on the number of infections, recoveries and deaths of 9 countries between 1/22/2020 and 4/3/2020, including Italy, Spain, Germany, France, the UK, Singapore, South Korea, China, and the United States. The dataset is made publicly available by the Center for Systems Science and Engineering (CSSE) at John Hopkins University and it can be downloaded at~\cite{JHU_COVID19_data}. In total, there are 73 days in this time span and the daily active cases (i.e., number of current infections) are plotted in Figure~\ref{fig:observed_data_COVID-19}.

In addition, we reviewed the NPIs each country imposed, and we considered the following NPIs in our initial DP-SIR model: travel restriction (TR), mask wearing (MW), lockdown (LD), social distancing (SD), school closure (SC), and centralized quarantine (CQ). 
The intervention times and descriptions of the NPIs imposed in each country were collected from local government websites, official public health authorities, and major newspapers~\cite{china_TR,china_LD,china_SD,china_SC,china_MW,china_CQ,sk_TR,sk_SC,sk_SD,sk_MW,sk_CQ,sg_TR,sg_SD,sg_CQ,it_TR,it_LD,it_SD,it_SC,esp_TR,esp_LD,esp_SD,esp_SC,fr_TR,fr_LD,fr_SD,fr_SC,ge_SD,ge_TR,ge_LD,ge_SD,ge_SC,uk_TR,uk_LD,uk_SD,uk_SC,us_TR,us_SD,us_SC}. Summary of each country's NPIs with their intervention times are shown in Table~\ref{tab:NPIs}. We choose the scale parameter $\tau = 7$ in~\eqref{eqn:model_NPIs} to control the time-lag effect of interventions for the COVID-19 study as a proxy for the incubation period (typically 2-14 days~\cite{incubation}).

\subsection{Estimation of NPI impact}

We first include all 6 NPIs from Table~\ref{tab:NPIs} into the DP-SIR model. The estimated coefficients with 95\% confidence intervals in this full model are shown in Table~\ref{tab:NPIs_estimate}.

\begin{table}[h!]
\begin{center}
\caption{NPIs used in our model. TR: travel restrictions; MW: mask wearing; LD: lockdown; SD: social distancing; SC: school closure; CQ: centralized quarantine. N/A means that an intervention is not implemented up to 4/3/20.}
\label{tab:NPIs}
\begin{tabular}{|l|c|c|c|c|c|c|}
\hline
County / NPI & TR & MW & LD & SD & SC & CQ \\ 
\hline\hline
Italy & 1/31/20 & N/A & 3/8/20 & 3/8/20 & 3/5/20 & N/A \\ \hline
Spain & 3/16/20 & N/A & 3/14/20 & 3/14/20 & 3/12/20 & N/A \\ \hline
Germany & 3/15/20 & N/A & 3/22/20 & 3/12/20 & 3/13/20 & N/A \\ \hline
France & 3/17/20 & N/A & 3.17/20 & 3/15/20 & 3/16/20 & N/A \\ \hline
UK & 3/17/20 & N/A & 3/23/20 & 3/16/20 & 3/20/20 & N/A \\ \hline
Singapore & 1/23/20 & N/A & N/A & 3/27/20 & N/A & 1/23/20 \setfootnotemark\label{Singapore}  \\ \hline
South Korea & 2/4/20 & 2/2/20 \setfootnotemark\label{SKorea}  & N/A & 3/20/20 & 2/24/20 & 3/2/20 \\ \hline
China & 1/23/20 & 2/2/20 \setfootnotemark\label{China} & 1/23/20 & 1/23/20 & 1/23/20 & 2/3/20 \\ \hline
U.S. & 2/2/20 & N/A & N/A & 3/16/20 & 3/16/20 & N/A \\ \hline
\end{tabular}
\end{center}
\end{table}

\afterpage{\footnotetext[\getrefnumber{Singapore}]{Singapore did not have a reported date for centralized quarantine. We used the earliest date for NPI as the proxy for central quarantine date.}
\footnotetext[\getrefnumber{SKorea}]{South Korea did not have a strict policy implementation date for wearing masks. We used 2/2/20 as the proxy date for wearing mask.} 
\footnotetext[\getrefnumber{China}]{Although mask wearing policy was imposed on 1/23/20, China had a shortage of mask supply until 2/2/20. BBC News Feburary 6th, 2020.}}

\begin{table}[h!]
\begin{center}
\caption{Estimated NPI impact.}
\label{tab:NPIs_estimate}
\begin{tabular}{|l|c|c|}
\hline
NPI & estimated coefficient & 95\% CI \\ 
\hline\hline
travel restriction (TR) & -0.343 & [-0.786, 0.100] \\
mask wearing (MW) & 0.651 & [0.009, 1.294] \\
lockdown (LD) & 1.063 & [0.427, 1.699] \\
social distancing (SD) & -0.279 & [-0.986, 0.427] \\
school closure (SC) & 0.972 & [0.339, 1.604] \\
centralized quarantine (CQ) & 2.042 & [1.493, 2.592] \\
\hline
\end{tabular}
\end{center}
\end{table}

Note that TR and SD are not statistically significant at the 95\% confidence level. First, while many countries imposed travel restrictions for passengers coming directly from China, they did not ban travels from other international destinations, which may not be an effective policy as people who were infected could still come across the border by connecting to a third country. Second, given that SD intervention came very close to SC and LD interventions, we may view SD as a weaker intervention than LD, while it is stronger than SC. Hence the inclusion of SC and LD would be a strong proxy for SD. In addition, the coefficients for MW, LD, SC, CQ are all positive, which implies that these interventions are effective in reducing COVID-19 transmission rate. Table~\ref{tab:NPIs_estimate} also indicates that CQ is the most effective NPI to mitigate the COVID-19 transmission, while LD, SC and MW also play an important role. In our subsequent scenario analyses, we primarily focus on the 4 NPIs with statistically significant positive coefficients (i.e., MW, LD, SC, CQ).

While NPIs have different impacts on mitigating virus transmission, they are also associated with different degrees of costs to the overall economy. Policies such as lockdown may hurt the economy significantly by forcing non-essential businesses to close for a certain period of time, which may cause firms to go bankruptcy and employees to lose jobs. Other policy such as mask wearing is the most cost-efficient due to the low cost of production and easy implementation. Understanding the benefits of different combinations of NPIs in reducing the transmission rates as well as the costs to the economy are essential for countries to choose the most appropriate combination of NPIs to balance the control of virus transmission as well as the economy performance. 

In the next section, we predict the COVID-19 transmission dynamics under various policy scenarios with different combinations of NPIs.

\subsection{Scenario analysis}

\begin{figure}[h] 
\centering
\caption{Predicted active cases of the 9 countries using MW+LD+SC+CQ. Date of the blue vertical line is 4/3/20. Observed number of active cases between 1/22/20 and 4/3/20 are on the left side of the blue vertical line and predicted number of active cases between 4/4/20 and 8/9/20 are on the right side of the blue vertical line.}
\label{fig:pred_active_cases_NPI_MW_LD_SC_CH}
\includegraphics[scale=0.7]{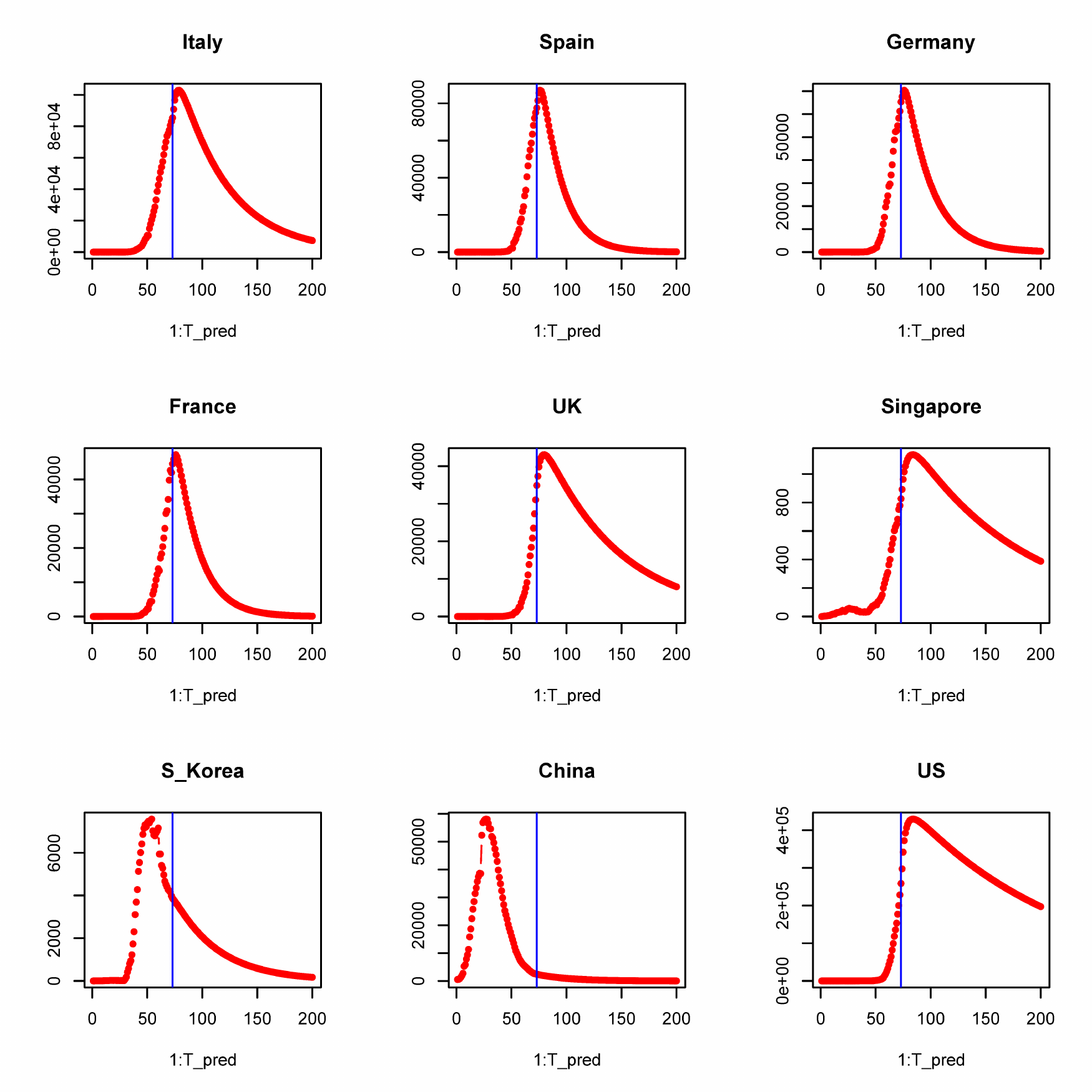}
\end{figure}

To begin with, we look at the predicted active cases of the 9 countries using the strongest combination MW+LD+SC+CQ, which also has the biggest negative impact on the economy. Predicted active cases of the 9 countries for MW+LD+SC+CQ over time up to 8/9/20 (i.e., $T = 200$) is shown in Figure~\ref{fig:pred_active_cases_NPI_MW_LD_SC_CH}. Predicted time point and height of the epidemic peak, as well as the number of total infected cases on 8/9/20 for MW+LD+SC+CQ are shown in Table~\ref{tab:MW+LD+SC+CQ_NPIs_pred}.

\begin{table}[h!]
\begin{center}
\caption{Predicted time point and height of the epidemic peak, as well as the number of total infected cases at $T = 200$ (i.e., 8/9/20) for MW+LD+SC+CQ.}
\label{tab:MW+LD+SC+CQ_NPIs_pred}
\begin{tabular}{|l|r|r|r|r|}
\hline
NPIs & Peak location & Peak value & Total \# cases by 8/9/20 & Total population \\ 
\hline\hline
Italy & 4/10/20 & 102,938 & 194,473 & 60,461,826 \\
Spain & 4/7/20 & 87,049 & 175,802 & 46,754,778 \\
Germany & 4/7/20 & 70,389 &124,911 & 83,783,942 \\
France & 4/7/20 & 47,219 & 88,022 & 65,273,511 \\
UK & 4/11/20 & 43,067 & 67,644 & 67,886,011 \\
Singapore & 4/15/20 & 1,136 & 2,846 & 5,850,342 \\
South Korea & 3/16/20 & 7577 & 11,013 & 51,269,185 \\
China & 2/18/20 & 58,108 & 83,073 & 1,439,323,776 \\
U.S. & 4/15/20 & 429,641 & 677,801 & 331,002,651 \\
\hline
\end{tabular}
\end{center}
\end{table}

\begin{table}[h!]
\begin{center}
\caption{Predicted time point and height of the epidemic peak, as well as the number of total infected cases at $T = 200$ (i.e., 8/9/20) for MW+SC+CQ.}
\label{tab:MW+SC+CQ_NPIs_pred}
\begin{tabular}{|l|r|r|r|r|}
\hline
NPIs & Peak location & Peak value & Total \# cases by 8/9/20 & Total population \\ 
\hline\hline
Italy & 4/11/20 & 113,933 & 233,740 & 60,461,826 \\
Spain & 4/8/20 & 95,805 & 203,657 & 46,754,778 \\
Germany & 4/8/20 & 75,893 & 144,758 & 83,783,942 \\
France & 4/8/20 & 50,414 & 98,814 & 65,273,511 \\
UK & 4/13/20 & 47,938 & 86,516 & 67,886,011 \\
Singapore & 4/18/20 & 1,201 & 3,647 & 5,850,342 \\
South Korea & 3/16/20 & 7,577 & 11,894 & 51,269,185 \\
China & 2/18/20 & 58,108 & 83,126 & 1,439,323,776 \\
U.S. & 4/19/20 & 389,914 & 774,269 & 331,002,651 \\
\hline
\end{tabular}
\end{center}
\end{table}

From Figure~\ref{fig:pred_active_cases_NPI_MW_LD_SC_CH}, we see that by imposing the strongest combination MW+LD+SC+CQ of NPIs, each country can reach the turning point the soonest, with Spain, Germany and France having the peak day on April 7th, Italy on April 10th, UK on April 11th, U.S. and Singapore on April 15th. After passing the peak days, the numbers of actively infected cases will drop faster for France, Spain and Germany than for Italy, UK, U.S. and Singapore. The NPIs would not impact China and South Korea significantly given that they both have passed the peak and especially China's active cases are approaching to zero.

\begin{figure}[h] 
\centering
\caption{Effective reproduction numbers $\cR_{\mbox{eff}}$ of the 9 countries using MW+LD+SC+CQ between 4/3/20 and 8/9/20. Blue horizontal line corresponds to $\cR_{\mbox{eff}} = 1$.}
\label{fig:effreprodnum_NPI_MW_LD_SC_CH}
\includegraphics[scale=0.7]{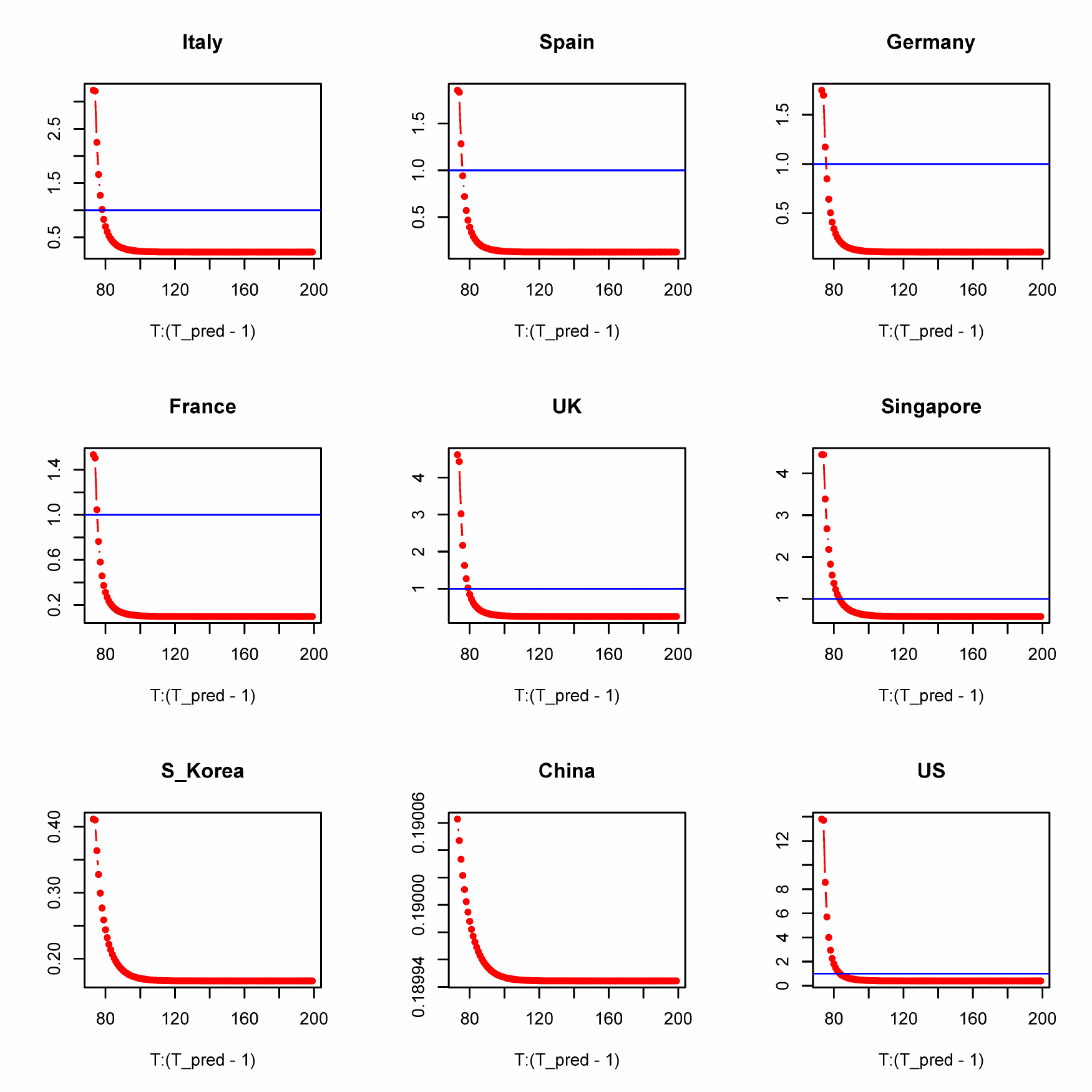}
\end{figure}

We notice that the MW+LD+SC+CQ scenario was indeed China's selection of NPIs during its outbreak. China took a strict implementation of all four policies, which succeeded in significantly reducing the reproduction rate in a timely manner (cf. Figure~\ref{fig:effreprodnum_NPI_MW_LD_SC_CH}). Although it is the most economically costly method in the short run, by getting the best infection control outcome in the shortest period of time, China can quickly reach a very low reproduction rate and recover the economy by releasing those strict policies sooner. Our finding is consistent with what China is doing right now: China lifts all those restrictions nationally, with Wuhan being the last city and planning to remove those policies on April 8th.

\begin{figure}[h] 
\centering
\caption{Predicted active cases of the 9 countries using MW+SC+CQ. Date of the blue vertical line is 4/3/20. Observed number of active cases between 1/22/20 and 4/3/20 are on the left side of the blue vertical line and predicted number of active cases between 4/4/20 and 8/9/20 are on the right side of the blue vertical line.}
\label{fig:pred_active_cases_NPI_MW_SC_CH}
\includegraphics[scale=0.7]{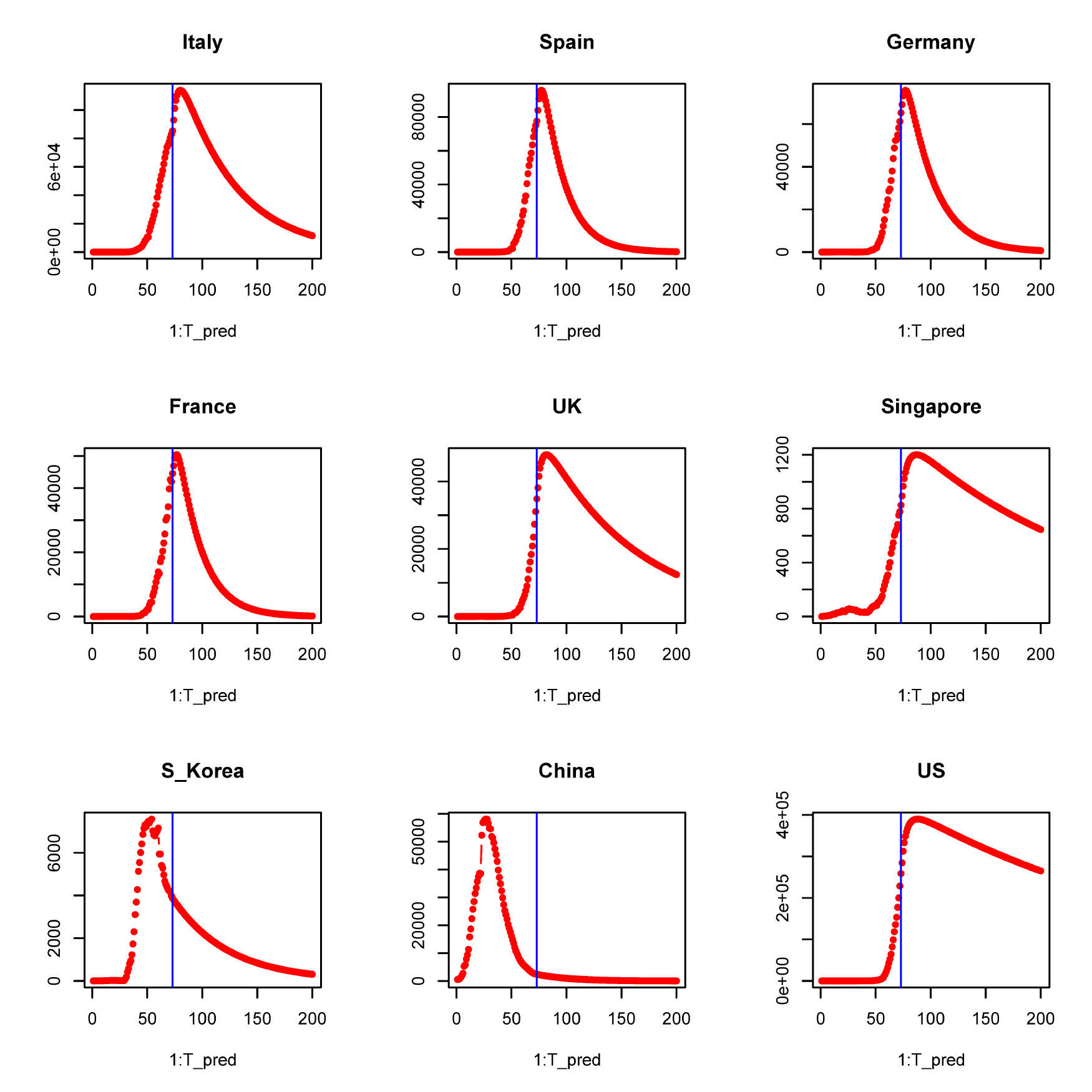}
\end{figure}

While having most strict NPIs can help each country which are experiencing outbreak to control virus transmission in the fastest way, the economy costs associated with such scenario are also the highest. Given the lockdown policy is in particular harmful to the economy, we then construct a weaker NPI combination by relaxing the lockdown policy while still imposing the other three: mask wearing, school closure, and centralized quarantine. Such NPI combination MW+SC+CQ can be economically affordable given it does not require closing businesses for all industries except for schools. The prediction is shown in Figure~\ref{fig:pred_active_cases_NPI_MW_SC_CH}.

By comparing the results of MW+SC+CQ in Figure~\ref{fig:pred_active_cases_NPI_MW_SC_CH} with MW+LD+SC+CQ in Figure~\ref{fig:effreprodnum_NPI_MW_LD_SC_CH}, we find that the predicted infection outcomes are very similar, which implies that lifting the lockdown policy may not substantially impact the infection outcome, provided that: (i) schools still remain closed, (ii) everyone is required to wear masks, and (iii) there is a centralized quarantine system to isolate all confirmed cases from their households. By excluding lockdown policy, the peak time is postponed for 1 day for Italy (April 11th), Spain, Germany and France (April 8th), 2 days for the UK (April 13th), 3 days for Singapore (April 18th) and 4 days for the U.S. (April 19th). Considering the significant economy harm of lockdown, this set of NPIs may be more feasible for many countries given it does not require to suspend the majority of economic activities on a national scale.

South Korea indeed took this approach MW+SC+CQ (without a national lockdown). By choosing this combination of NPIs, South Korea reached its turning point fairly quickly and successfully managed to control the COVID-19 transmission in an economically efficient way. Although it may control the virus spread slower and may require lifting those imposed policies later than what China did, it does provide a solution to keep the entire economy running even during the outbreak. This finding may provide insights for European countries, which are experiencing the negative economic consequences caused by the national lockdown, to consider imposing mask wearing and centralized quarantine so they can lift the lockdown policy. 

To proceed further, we remove one more NPI from MW+SC+CQ to see whether or not any combination of two NPIs can achieve as efficient outcomes as imposing all the three, which may be more economically efficient. While imposing two NPIs generally work for majority of the European countries (see the complete data in Appendix), we find that it may not work effectively for Singapore and the U.S. In particular, we find that by imposing only SC+CQ or MW+CQ, Singapore cannot even reach the turning points by August 9th, which implies that the transmission rate will not be below 1 and the outbreak cannot be controlled. For the U.S., although imposing two NPIs can still enable the country to reach the turning points in late April (April 25th for SC+CQ, April 25th for MW+SC, and April 22nd for MW+CQ), but after passing the turning point, the transmission rates drop very slowly, with the curve for active cases remain relative flat. It suggests that both Singapore and the U.S. may need to impose stricter NPIs relative to other countries in order to enable reaching a turning point and/or faster dropping transmission rates. 


To summarize, our scenario analysis provides useful insights for countries to choose the most appropriate combination of NPIs which can balance the benefits on infection controls and the costs to the overall economy. Taking into consideration the heterogeneity across countries, each country should customize its own choices of NPIs to reach its desired goal in both infection and economy measures. As of April 4th, Singapore and the U.S. may need to impose stricter NPIs, while European countries which are about to reach the turning points in the next few days can cautiously consider to gradually lift some policy such as national lockdown. The estimation and scenario analysis results provide an empirical evidence for countries to impose mask wearing and centralized quarantines instead of a national lockdown, which can provide similar disease transmission-control outcomes and meanwhile minimize the damage to the overall economy. Our findings can also provide useful insights for other countries which have not yet and however may experience an epidemic outbreak in a future time. 

\section{Discussion}
\label{sec:discussion}

This paper proposes a dynamic SIR model to estimate the impact of various NPIs on the COVID-19 transmission using panel data from 9 countries across the globe. Data from these countries show significant variations in their selections and timelines of the NPIs. Our findings suggest that centralized quarantine is the most effective NPI measure, followed by lockdown, school closure and wearing masks. The scenario analysis shows that lockdown might be cautiously lifted if the country using the other three NPIs simultaneously (school closure, wearing masks and centralized quarantine). 

Our findings provide feasible solutions for countries to use economically affordable NPIs such as mask wearing and centralized quarantine to replace the highly economically costly NPIs such as national lockdown, without significantly heightening the epidemic peak and can substantially flatten the curve of the active COVID-19 cases to reach a non-epidemic regime. This paper also suggests each country should customize its choice of NPIs by considering specific socioeconomic situations within the country. In particular, our empirical findings suggest that, as of April 4th, Singapore and the U.S. might consider imposing stricter NPIs, while European countries may cautiously consider to gradually lift the policies after reaching the turning points in a few days.

We are aware that there are likely undetected cases in the population, for instance because the testing capacity is limited and individual has no symptoms, which may cause potential issues such as the second-wave of the epidemic after the control policy is lifted. It is possible to calibrate such effect by ongoing random testing of the population in the SIR framework~\cite{Stock2020}.


To conclude, the DP-SIR model used with panel data can provide useful insights for countries to choose the most appropriate NPIs in a timely manner to balance the desired goals between the economy performance and health consequence.


%

\bibliographystyle{plain}
\bibliography{dp-sir}

\clearpage
\newpage
\appendix

\section{Additional prediction results}

This appendix presents additional prediction results.

\begin{table}[h!]
\begin{center}
\caption{Predicted time point and height of the epidemic peak, as well as the number of total infected cases at $T = 200$ (i.e., 8/9/20) in Italy. Total population size of Italy is 60,461,826.}
\label{tab:Italy_various_NPIs}
\begin{tabular}{|l|r|r|r|}
\hline
NPIs & Peak location & Peak value & Total number of cases by 8/9/20 \\ 
\hline\hline
SC+CQ & 4/12/20 & 115,495 & 259,519 \\
MW+CQ & 4/12/20 & 153,948 & 287,300 \\
MW+SC & 4/12/20 & 110,310 & 258,353 \\
MW+SC+CQ & 4/11/20 & 113,933 & 233,740 \\
MW+LD+SC+CQ & 4/10/20 & 102,938 & 194,473 \\
\hline
\end{tabular}
\end{center}
\end{table}

\begin{table}[h!]
\begin{center}
\caption{Predicted time point and height of the epidemic peak, as well as the number of total infected cases at $T = 200$ (i.e., 8/9/20) in Spain. Total population size of Spain is 46,754,778.}
\label{tab:Spain_various_NPIs}
\begin{tabular}{|l|r|r|r|}
\hline
NPIs & Peak location & Peak value & Total number of cases by 8/9/20 \\ 
\hline\hline
SC+CQ & 4/9/20 & 95,918 & 214,319 \\
MW+CQ & 4/10/20 & 145,736 & 290,562 \\
MW+SC & 4/8/20 & 90,480 & 205,233 \\
MW+SC+CQ & 4/8/20 & 95,805 & 203,657 \\
MW+LD+SC+CQ & 4/7/20 & 87,049 & 175,802 \\
\hline
\end{tabular}
\end{center}
\end{table}

\begin{table}[h!]
\begin{center}
\caption{Predicted time point and height of the epidemic peak, as well as the number of total infected cases at $T = 200$ (i.e., 8/9/20) in Germany. Total population size of Germany is 83,783,942.}
\label{tab:Germany_various_NPIs}
\begin{tabular}{|l|r|r|r|}
\hline
NPIs & Peak location & Peak value & Total number of cases by 8/9/20 \\ 
\hline\hline
SC+CQ & 4/8/20 & 75,799 & 151,653 \\
MW+CQ & 4/10/20 & 110,708 & 205,406 \\
MW+SC & 4/8/20 & 72,260 & 145,375 \\
MW+SC+CQ & 4/8/20 & 75,893 & 144,758 \\
MW+LD+SC+CQ & 4/7/20 & 70,389 &124,911 \\
\hline
\end{tabular}
\end{center}
\end{table}

\begin{table}[h!]
\begin{center}
\caption{Predicted time point and height of the epidemic peak, as well as the number of total infected cases at $T = 200$ (i.e., 8/9/20) in France. Total population size of France is 65,273,511.}
\label{tab:France_various_NPIs}
\begin{tabular}{|l|r|r|r|}
\hline
NPIs & Peak location & Peak value & Total number of cases by 8/9/20 \\ 
\hline\hline
SC+CQ & 4/8/20 & 50,338 & 102,452 \\
MW+CQ & 4/10/20 & 74,161 & 141,862 \\
MW+SC & 4/7/20 & 48,061 & 97,952 \\
MW+SC+CQ & 4/8/20 & 50,414 & 98,814 \\
MW+LD+SC+CQ & 4/7/20 & 47,219 & 88,022 \\
\hline
\end{tabular}
\end{center}
\end{table}

\begin{table}[h!]
\begin{center}
\caption{Predicted time point and height of the epidemic peak, as well as the number of total infected cases at $T = 200$ (i.e., 8/9/20) in the UK. Total population size of the UK is 67,886,011.}
\label{tab:UK_various_NPIs}
\begin{tabular}{|l|r|r|r|}
\hline
NPIs & Peak location & Peak value & Total number of cases by 8/9/20 \\ 
\hline\hline
SC+CQ & 4/14/20 & 48,627 & 97,341 \\
MW+CQ & 4/15/20 & 77,132 & 146,285 \\
MW+SC & 4/14/20 & 45,537 & 91,453 \\
MW+SC+CQ & 4/13/20 & 47,938 & 86,516 \\
MW+LD+SC+CQ & 4/11/20 & 43,067 & 67,644 \\
\hline
\end{tabular}
\end{center}
\end{table}

\begin{table}[h!]
\begin{center}
\caption{Predicted time point and height of the epidemic peak, as well as the number of total infected cases at $T = 200$ (i.e., 8/9/20) in Singapore. N/A means peak is not achieved by 8/9/20. Total population size of Singapore is 5,850,342.}
\label{tab:Singapore_various_NPIs}
\begin{tabular}{|l|r|r|r|}
\hline
NPIs & Peak location & Peak value & Total number of cases by 8/9/20 \\ 
\hline\hline
SC+CQ & N/A & 3,239 & 9,587 \\
MW+CQ & N/A & 10,290 & 22,792 \\
MW+SC & 4/10/20 & 1,121 & 1,861 \\
MW+SC+CQ & 4/18/20 & 1,201 & 3,647 \\
MW+LD+SC+CQ & 4/15/20 & 1,136 & 2,846 \\
\hline
\end{tabular}
\end{center}
\end{table}

\begin{table}[h!]
\begin{center}
\caption{Predicted time point and height of the epidemic peak, as well as the number of total infected cases at $T = 200$ (i.e., 8/9/20) in the U.S. Total population size of the U.S. is 331,002,651.}
\label{tab:US_various_NPIs}
\begin{tabular}{|l|r|r|r|}
\hline
NPIs & Peak location & Peak value & Total number of cases by 8/9/20 \\ 
\hline\hline
SC+CQ & 4/25/20 & 404,489 & 922,658 \\
MW+CQ & 4/22/20 & 641,340 & 1,315,976 \\
MW+SC & 4/25/20 & 378,099 & 874,759 \\
MW+SC+CQ & 4/19/20 & 389,914 & 774,269 \\
MW+LD+SC+CQ & 4/15/20 & 429,641 & 677,801 \\
\hline
\end{tabular}
\end{center}
\end{table}



\begin{figure}[h] 
\centering
\caption{Predicted active cases of the 9 countries using MW+SC. Date of the blue vertical line is 4/3/20. Observed number of active cases between 1/22/20 and 4/3/20 are on the left side of the blue vertical line and predicted number of active cases between 4/4/20 and 8/9/20 are on the right side of the blue vertical line.}
\label{fig:pred_active_cases_NPI_MW_SC}
\includegraphics[scale=0.7]{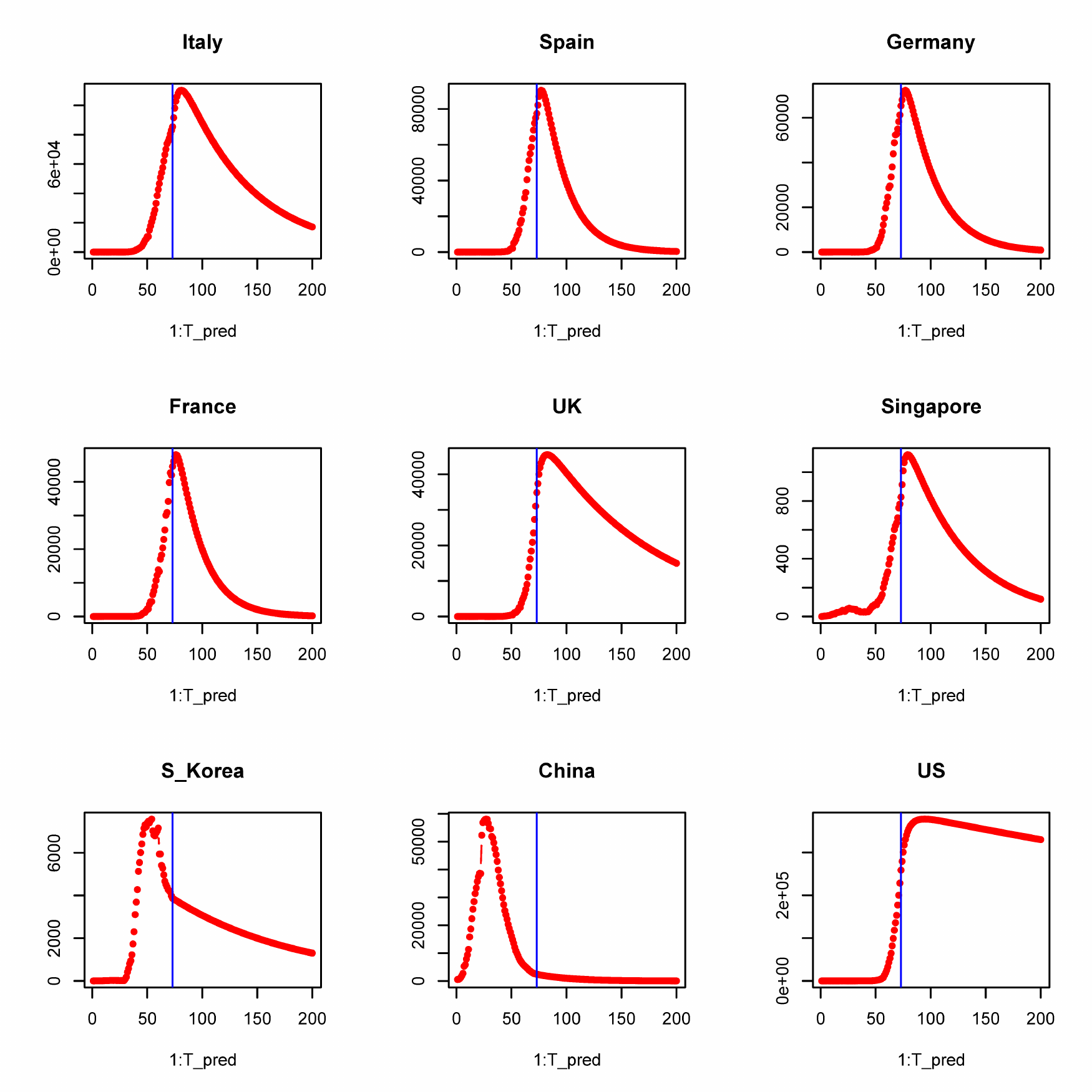}
\end{figure}

\begin{figure}[h] 
\centering
\caption{Predicted active cases of the 9 countries using MW+CQ. Date of the blue vertical line is 4/3/20. Observed number of active cases between 1/22/20 and 4/3/20 are on the left side of the blue vertical line and predicted number of active cases between 4/4/20 and 8/9/20 are on the right side of the blue vertical line.}
\label{fig:pred_active_cases_NPI_MW_CH}
\includegraphics[scale=0.7]{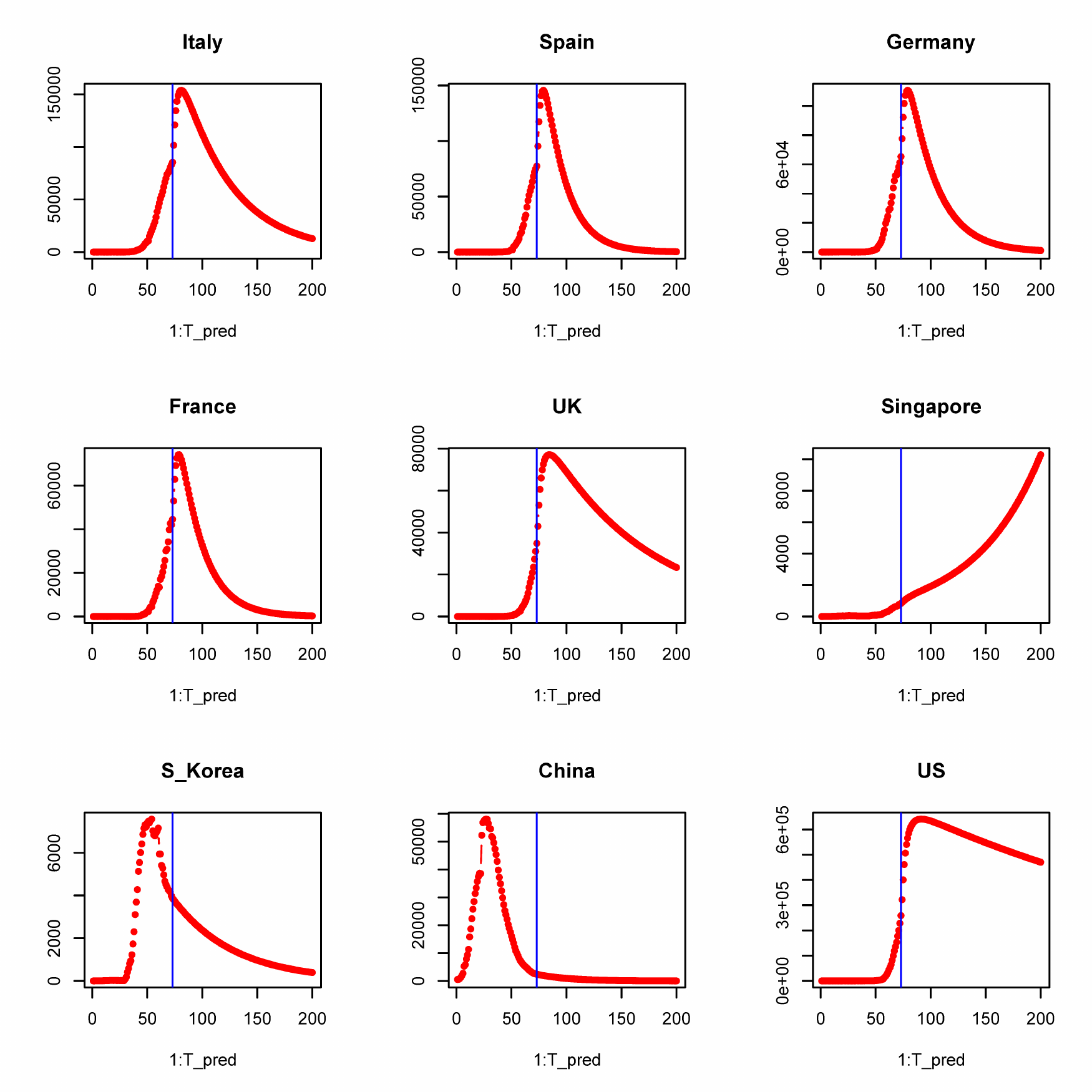}
\end{figure}


\begin{figure}[h] 
\centering
\caption{Predicted active cases of the 9 countries using SC+CQ. Date of the blue vertical line is 4/3/20. Observed number of active cases between 1/22/20 and 4/3/20 are on the left side of the blue vertical line and predicted number of active cases between 4/4/20 and 8/9/20 are on the right side of the blue vertical line.}
\label{fig:pred_active_cases_NPI_SC_CH}
\includegraphics[scale=0.7]{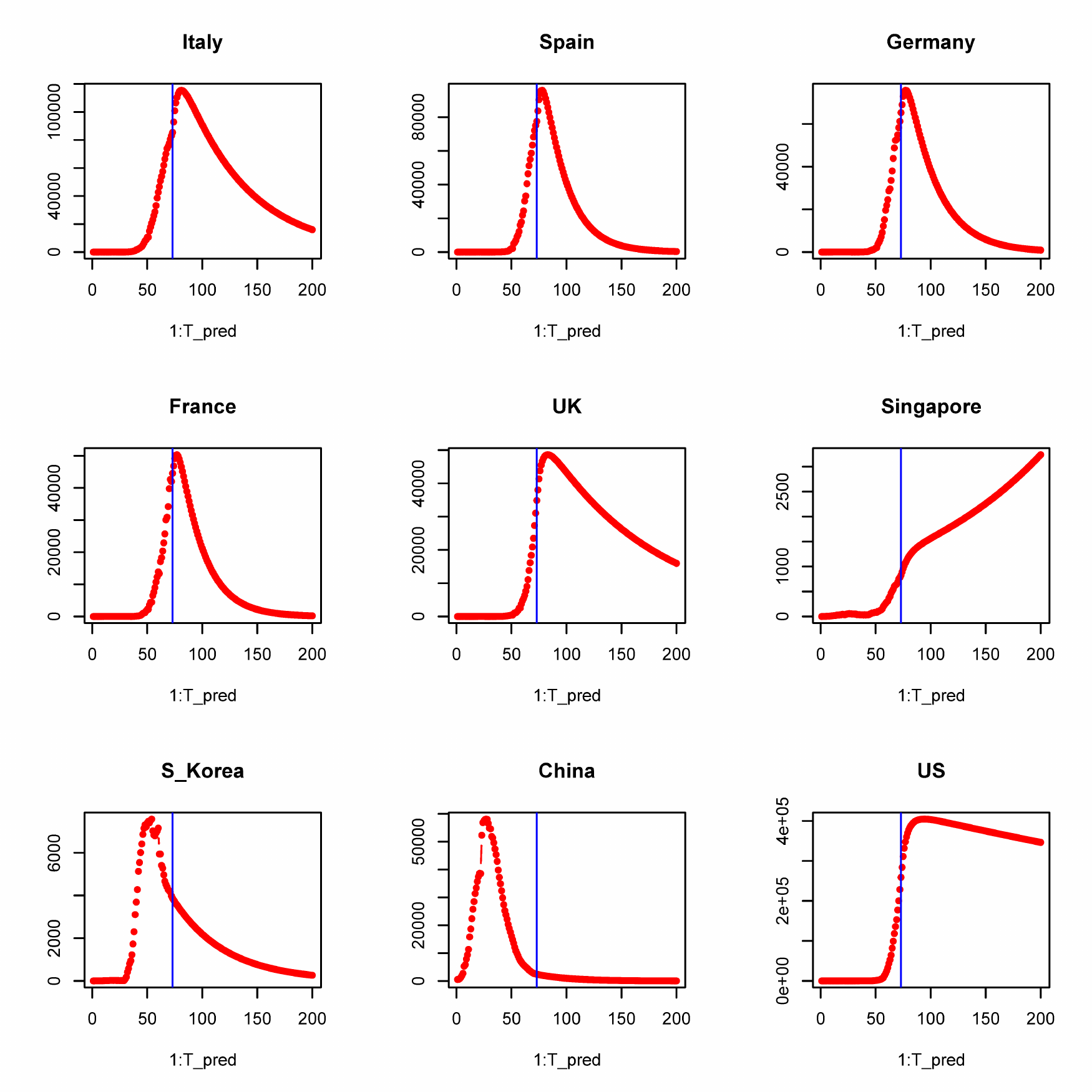}
\end{figure}

\begin{figure}[h] 
\centering
\caption{Effective reproduction numbers $\cR_{\mbox{eff}}$ of the 9 countries using MW+SC+CQ between 4/3/20 and 8/9/20. Blue horizontal line corresponds to $\cR_{\mbox{eff}} = 1$.}
\label{fig:effreprodnum_NPI_MW_SC_CH}
\includegraphics[scale=0.7]{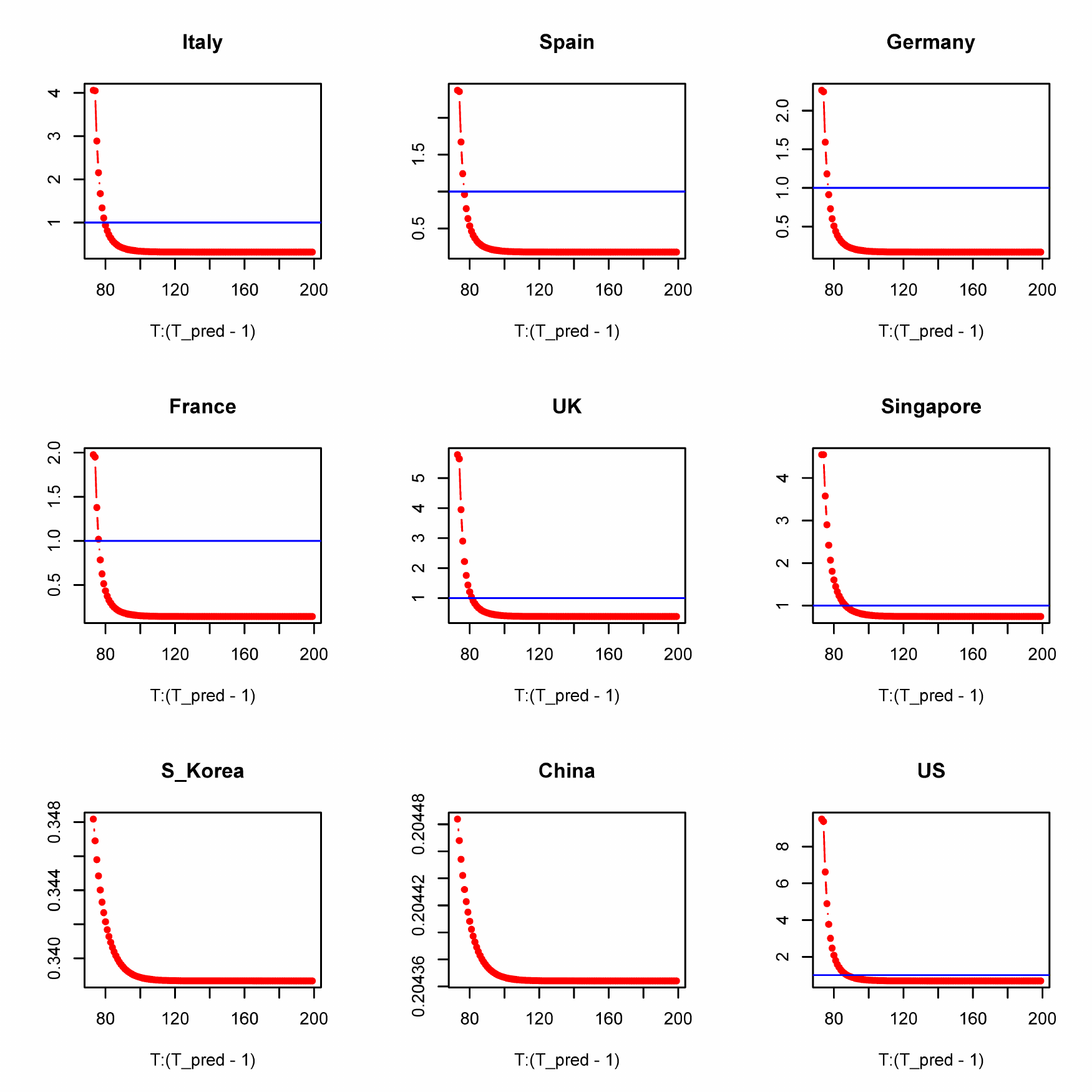}
\end{figure}



\begin{figure}[h] 
\centering
\caption{Effective reproduction numbers $\cR_{\mbox{eff}}$ of the 9 countries using MW+SC between 4/3/20 and 8/9/20. Blue horizontal line corresponds to $\cR_{\mbox{eff}} = 1$.}
\label{fig:effreprodnum_NPI_MW_SC}
\includegraphics[scale=0.7]{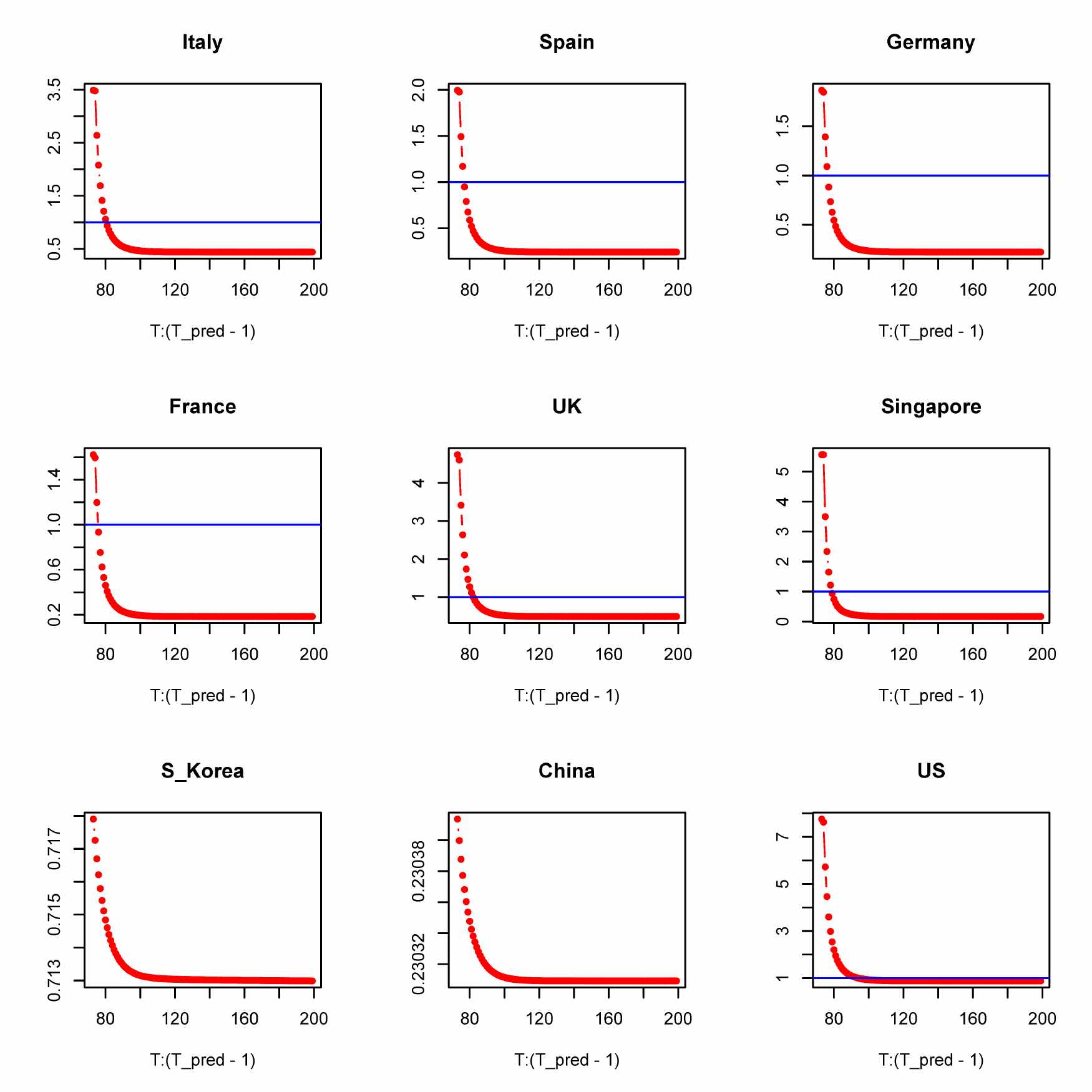}
\end{figure}

\begin{figure}[h] 
\centering
\caption{Effective reproduction numbers $\cR_{\mbox{eff}}$ of the 9 countries using MW+CQ between 4/3/20 and 8/9/20. Blue horizontal line corresponds to $\cR_{\mbox{eff}} = 1$.}
\label{fig:effreprodnum_NPI_MW_CH}
\includegraphics[scale=0.7]{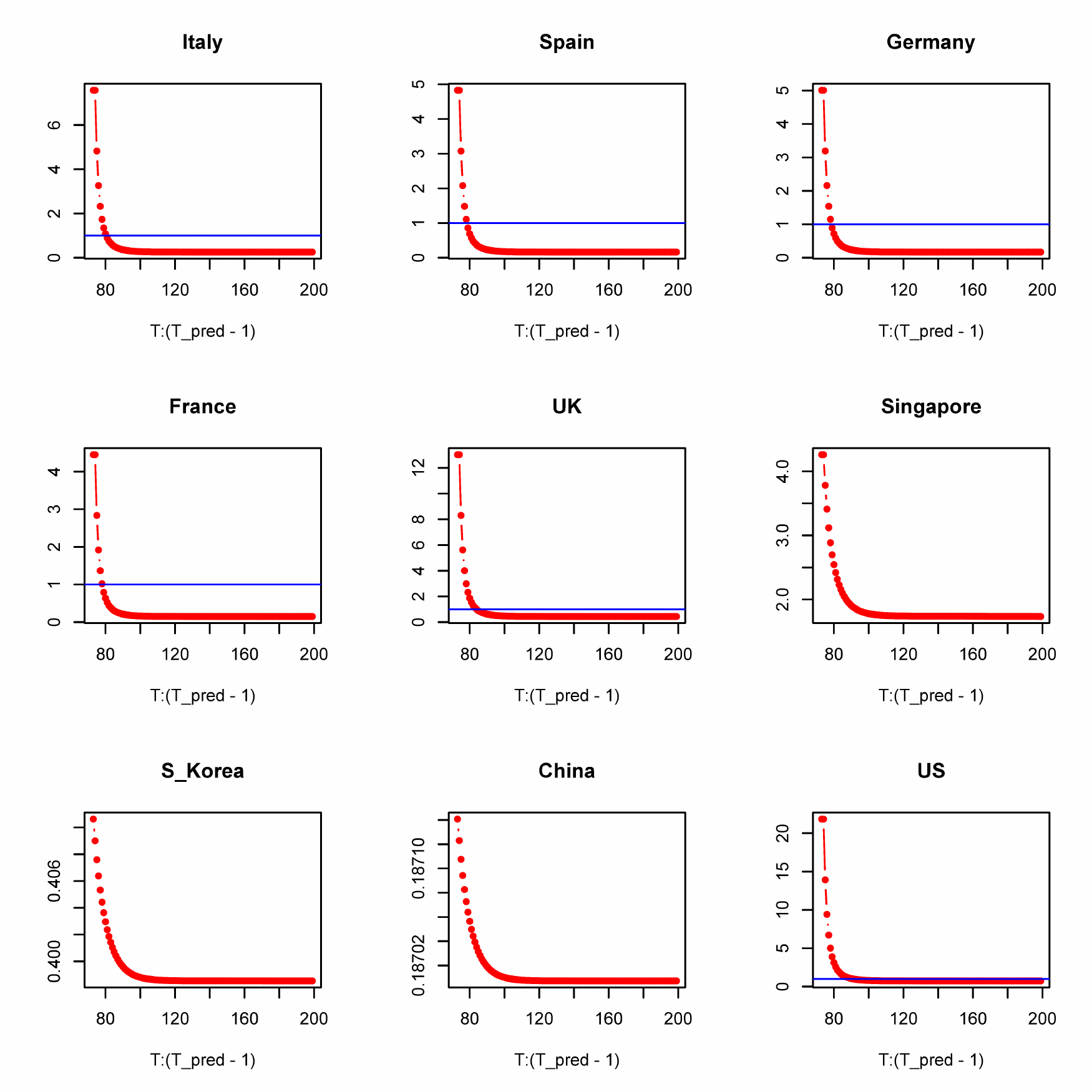}
\end{figure}


\begin{figure}[h] 
\centering
\caption{Effective reproduction numbers $\cR_{\mbox{eff}}$ of the 9 countries using SC+CQ between 4/3/20 and 8/9/20. Blue horizontal line corresponds to $\cR_{\mbox{eff}} = 1$.}
\label{fig:effreprodnum_NPI_SC_CH}
\includegraphics[scale=0.7]{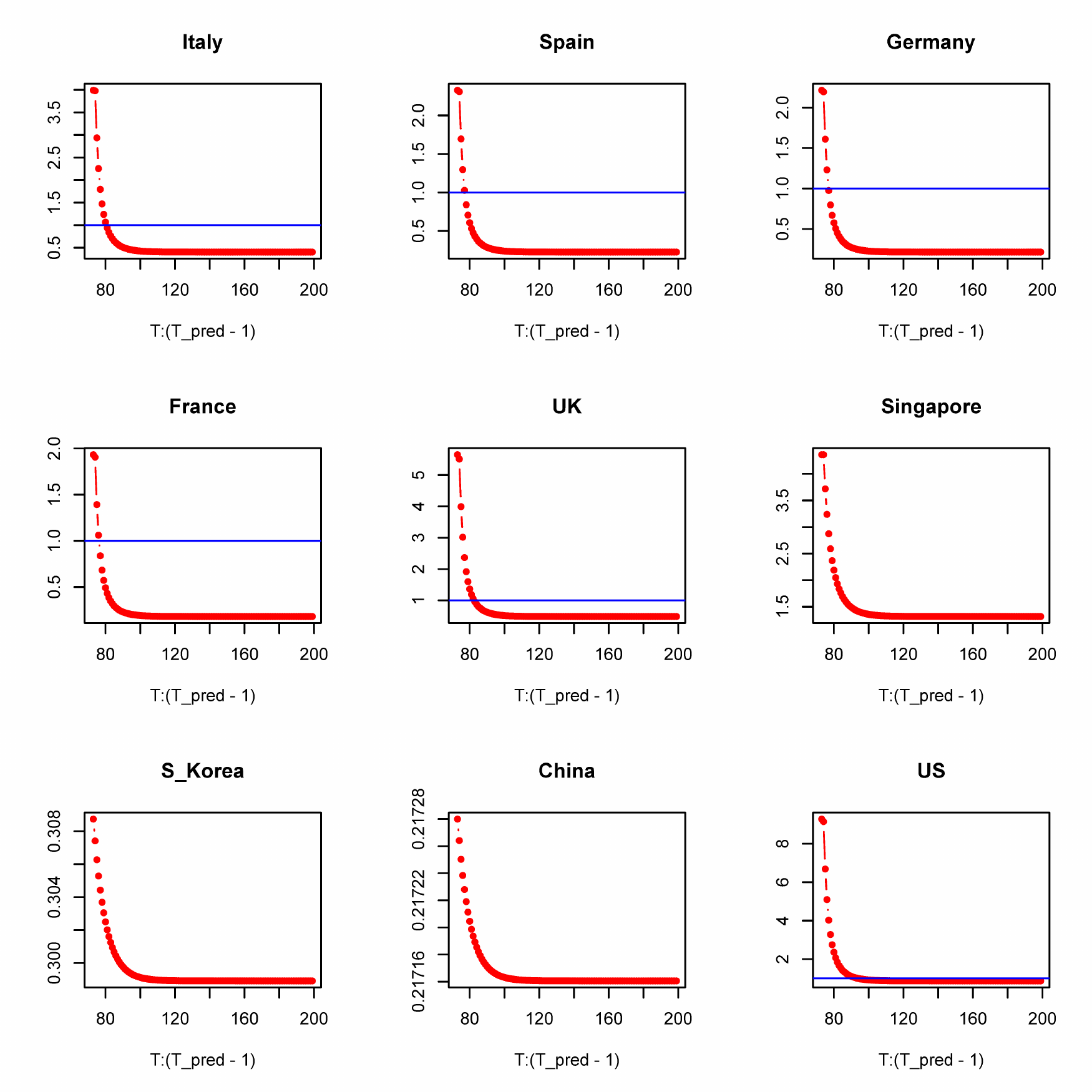}
\end{figure}

\end{document}